\documentclass[11pt]{article}
\usepackage{moriond,epsfig}

\newcommand{\Dphi}{\Delta \phi\,{}_{\mathrm{dijet}}}
\newcommand{\ptmax}{p_T^{{\mathrm{max}}}}

\begin{document}
\vspace*{4cm}
\title{\boldmath  
Latest Jets Results from the Tevatron at $\sqrt{s}=1.96\,\mathrm{TeV}$ }
\author{
Alexander Kup\v{c}o
\\
\vskip 0.30cm                                                                 
\centerline{(on behalf of the CDF and D\O\ Collaborations)}
\vskip 0.30cm
}
\address{
\centerline{Institute of Physics, Academy of Sciences, Center          
           for Particle Physics, Prague, Czech Republic}
}
\maketitle\abstracts{
Latest jet results from the Tevatron
are presented in this conference note. These are namely:
new results on central inclusive jet production
using both cone and $k_T$ algorithms, measurement of decorrelation
in azimuthal angle between the two jets with the highest transverse
momenta,
and study of jet shapes.
Results are based on data
collected in $\mathrm{p}\bar{\mathrm{p}}$ collisions at $\sqrt{s}=1.96\,\mathrm{TeV}$  in the
 years 2001-2004.  Depending on the analysis, integrated 
luminosity of the sample was up to $378\,\mathrm{pb}^{-1}$.
}

Producing events with high transverse momenta 
($p_{\mathrm{T}}$) jets we probe properties of matter and space at 
very short distances. At the Fermilab Tevatron, jets are being produced  with 
$p_{\mathrm{T}}$ up to about $600\,\mathrm{GeV}$ which corresponds to 
distances about thousand times smaller than the proton size. 
Up to this scale, the
Tevatron high $p_{\mathrm{T}}$ jet data provide
tests of our understanding of proton structure and strong force
that acts between proton constituents: quarks and gluons.
Another aspect of the Tevatron jet physics program is the detailed study
of jet properties and understanding these properties within the
framework of perturbative Quantum Chromodynamics (pQCD). 
These studies have direct impact on many non-QCD
analyzes that work with jets. In addition, strong interaction induced 
processes provide unavoidable background for these analyzes in 
hadron-hadron colliders. Better understanding of QCD processes
thus improve our chances for discovery of potential signals of new physics.

To the first category of high $p_{\mathrm{T}}$ jet program belong
updated results on central jet inclusive $p_{\mathrm{T}}$ 
spectra both from the CDF and the D\O\ Collaborations. Second category is 
represented by the D\O\ measurement of dijet azimuthal decorrelations, 
study of jet shapes by CDF, and the CDF measurement of inclusive jet 
$p_{\mathrm{T}}$ spectra using $k_{\mathrm{T}}$ algorithm with
different sizes of jets.

The analyzes presented in this talk are based on data collected
with CDF~\cite{cdf} and D\O~\cite{d0det} detectors in Run~II of the Tevatron. 
This period started in year 2001 after upgrade of collider and both
detectors. The Tevatron delivers now proton antiproton collisions
at $\sqrt{s}=1.96\,\mathrm{TeV}$ while in previous run (Run~I) it was
$\sqrt{s}=1.8\,\mathrm{TeV}$. Also Tevatron luminosity was increased
significantly. Both aspects of collider upgrade affected
high $p_{\mathrm{T}}$ jet physics. Due to them, both experiments 
already collected about one order of magnitude more events 
with high $p_{\mathrm{T}}$ jets than in Run~I.

\begin{figure}
\begin{minipage}[t]{7.5cm}
\includegraphics[scale=0.31]{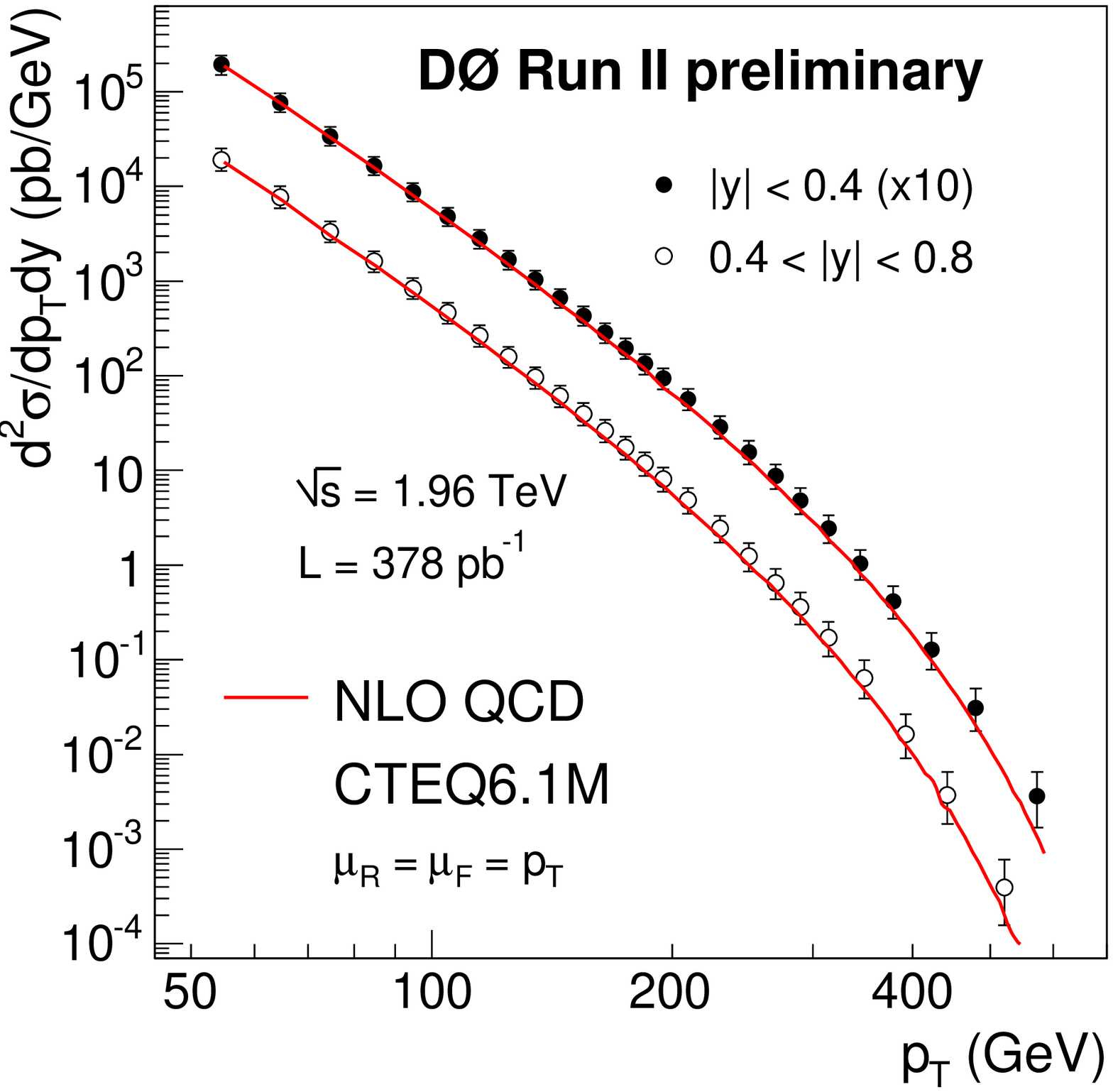}
\caption{\label{f:d0jetpt} 
Jet inclusive cross section measured by D\O. Error bars represent
total error (statistical and systematic errors added in quadrature).
For $|y|<0.4$ rapidity bin,
the results are scaled up by factor of 10.}
\end{minipage}
\hfill
\begin{minipage}[t]{7.5cm}
\includegraphics[scale=0.39]{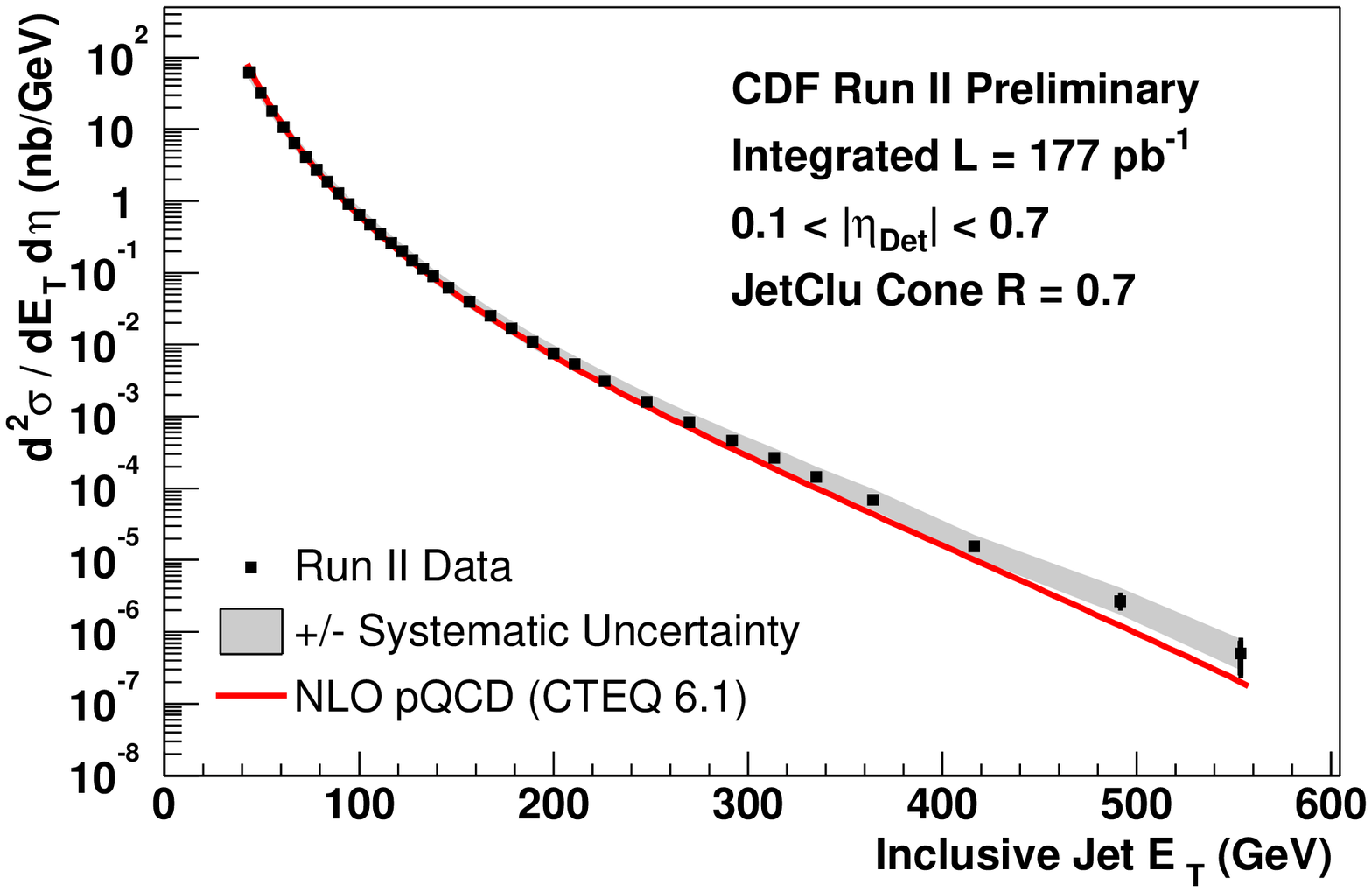}
\caption{\label{f:cdfjetpt} 
Jet inclusive cross section measured by CDF.}
\end{minipage}
\end{figure}

The CDF and D\O\ Collaborations measured inclusive production of central
jets. D\O\ performed the measurement in two rapidity
bins: $|y|<0.4$ and $0.4<|y|<0.8$ (rapidity $y$ is defined as
$y={1\over2}\ln{E+p_z\over E-p_z}$, where $E$ is jet energy and $p_z$ is jet
longitudinal momentum along the beam axis). The measurement was based
on $378\,\mathrm{pb}^{-1}$ of data. Jets were reconstructed using
iterative seed-based cone algorithm (including midpoints) with radius
$R=0.7$~\cite{run2cone}. Jet energies where calibrated using
$\gamma+\mathrm{jet}$ sample. Uncertainty on jet energy was found to be 
between 4-5\%,  for jet energies between $30$-$400\,\mathrm{GeV}$, 
and about 6\% at $500\mathrm{GeV}$. It translates to about 25\% (30\%, 60\%) 
error on jet production cross section for $p_{\mathrm{T}}=100\,\mathrm{GeV}$
($300\,\mathrm{GeV}$, $500\,\mathrm{GeV}$). Results of the measurement
are shown in Fig.~\ref{f:d0jetpt}.
The error on jet cross section was dominated by the uncertainty 
on jet energy calibration. 

CDF performed the measurement for jets with $0.1<|\eta|<0.7$ 
where $\eta$ is jet pseudorapidity ($\eta=-\ln\tan\vartheta/2$, where
$\vartheta$ is jet polar angle measured from the beam axis).
In this case, jets where reconstructed with CDF Run~I cone algorithm
with radius $R=0.7$~\cite{cdfRunIcone}. Measurement was based on
$177\,\mathrm{pb}^{-1}$ of data. Final result is shown in 
Fig.~\ref{f:cdfjetpt}. Systematic error is again dominated by the
uncertainty of jet energy calibration which was about 3\%.
CDF calibrated their jet energies using calorimeter electron
and hadronic responses measured during testbeam.

\begin{figure}
\begin{minipage}[t]{7.5cm}
\includegraphics[scale=0.4]{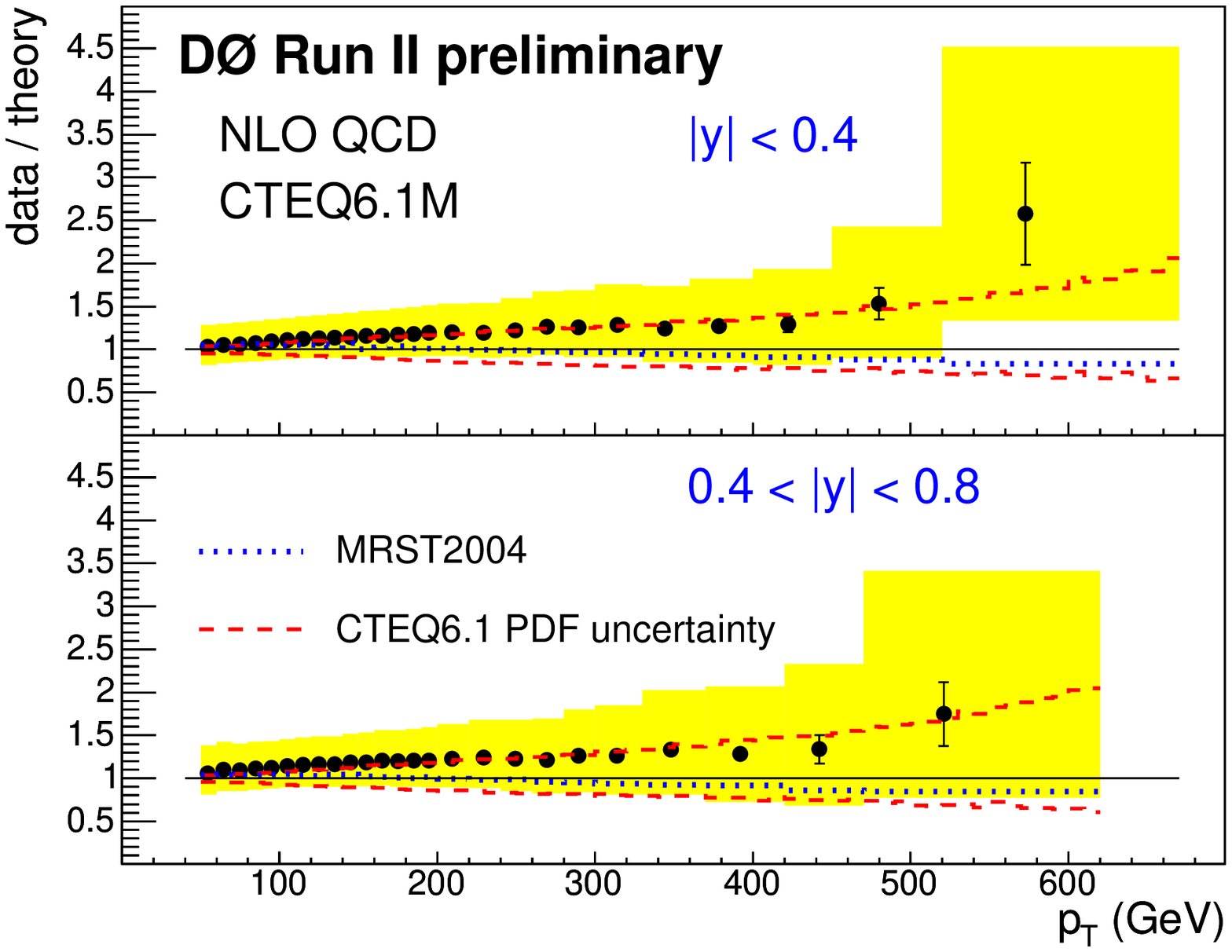}
\caption{\label{f:nlod0} 
Ratio of jet inclusive cross sections measured by D\O\ to NLO QCD predictions.
Filled regions represent total experimental error.
}
\end{minipage}
\hfill
\begin{minipage}[t]{7.5cm}
\includegraphics[scale=0.4]{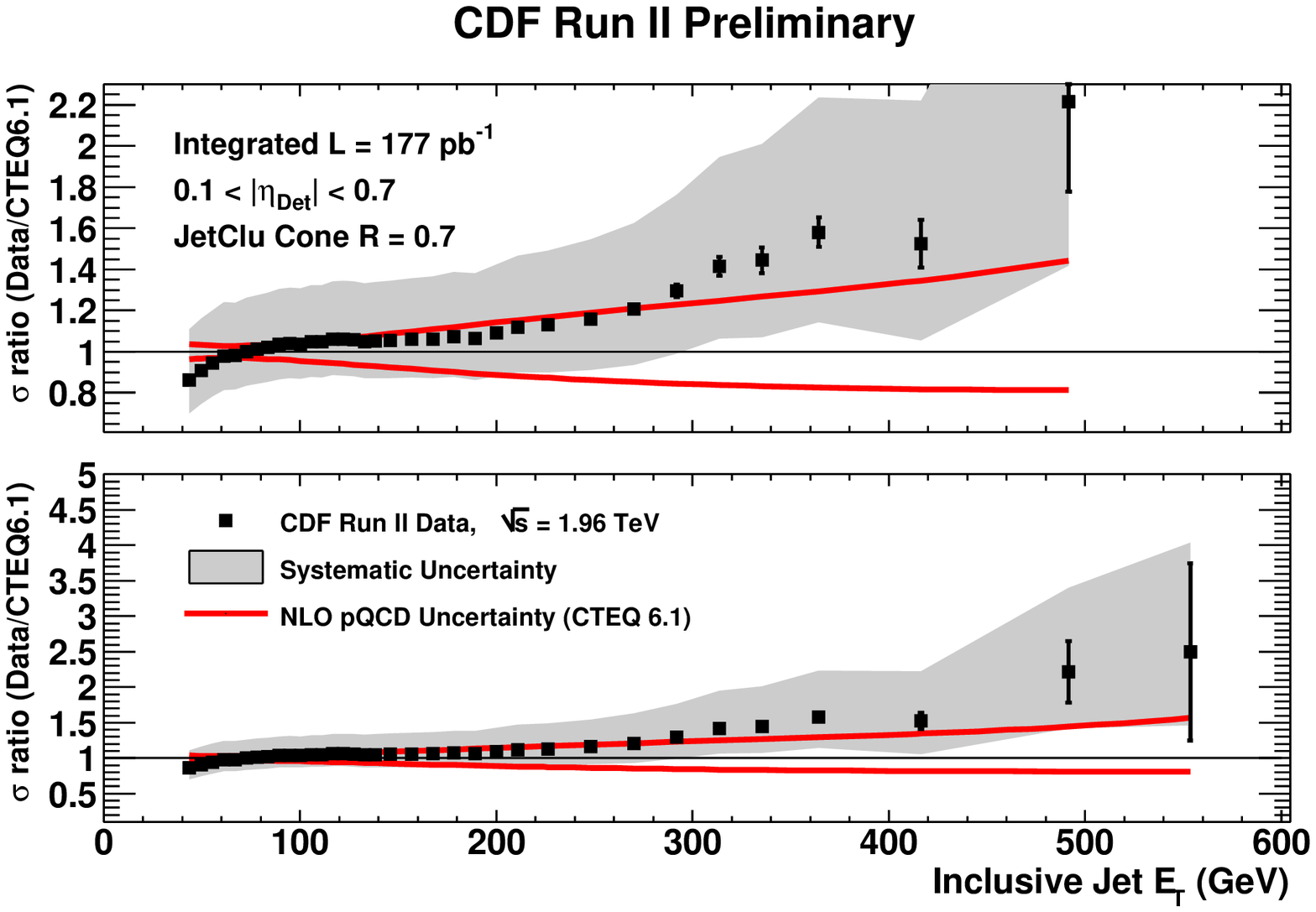}
\caption{\label{f:nlocdf} 
Ratio of jet inclusive cross section measured by CDF to NLO QCD prediction.
}
\end{minipage}
\end{figure}

CDF and D\O\ results were compared with next-to-leading (NLO) QCD 
predictions. In both cases, a good agreement was observed over
the entire region of jet $p_{\mathrm{T}}$ (from $50\,\mathrm{GeV}$
up to about $600\,\mathrm{GeV}$) in which the cross section is
rapidly falling down by 8 orders of magnitude. More detailed
comparison with NLO QCD prediction is given in Fig.~\ref{f:nlod0}
(D\O) and Fig.~\ref{f:nlocdf} (CDF). CTEQ6.1M~\cite{cteq6}
and MRST2004~\cite{mrst04} parton distribution functions (PDF) where
used in the NLO QCD calculations. Both sets of PDF lead to
similar predictions. 
Data are sensitive to running of strong coupling 
$\alpha_{\mathrm{S}}$ and also to proton structure functions.
At high $p_{\mathrm{T}}$, theoretical uncertainty
is dominated by the uncertainty
on gluon distribution function at high $x$ (where $x$ is fraction of proton momentum
carried by gluon). 

\begin{figure}
\begin{center}
\includegraphics[scale=0.9]{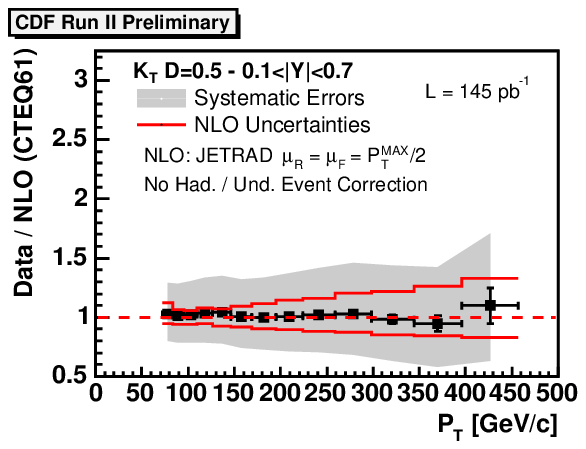}
\hspace{10mm}
\includegraphics[scale=0.9]{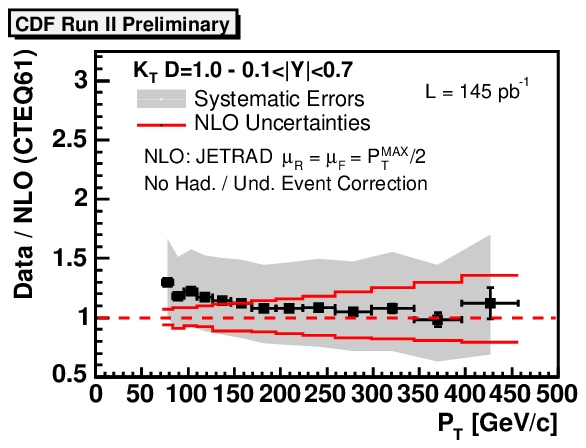}
\end{center}
\caption{\label{f:cdf_d05_10} 
Comparison of $k_{\mathrm{T}}$ jet inclusive cross sections with NLO QCD 
for $D=0.5$ (left) and $D=1.0$ (right). NLO QCD was calculated using
{\sc jetrad}$^{\,8}$.
}
\end{figure}

The CDF Collaboration also studied central jet production 
for jets reconstructed with $k_{\mathrm{T}}$-algorithm. 
CDF used
Ellis-Sopper version of $k_{\mathrm{T}}$-algorithm~\cite{ellis_soper} 
adapted for hadron-hadron colliders.
In this case, the et size is controlled by the parameter $D$.
Obtained jet $p_T$ cross section for $D=0.7$ and its comparison with 
NLO QCD predictions lead to the same conclusions as in the 
case of cone jets.
CDF performed the measurement for three different sizes of 
$k_{\mathrm{T}}$ jets ($D=0.5$, 0.7, and 1.0). For
$p_{\mathrm{T}}>150\,\mathrm{GeV}$, there was, 
with respect to NLO QCD, no difference between them.
At low $p_T$ end, observed
jet production is higher above the NLO QCD prediction for higher
values of $D$ (see Fig.~\ref{f:cdf_d05_10}).
The results suggest that the larger jets are more sensitive to
hadronization effects and/or to the soft underlying event physics.

\begin{figure}
\begin{minipage}[t]{7.5cm}
\includegraphics[scale=0.8]{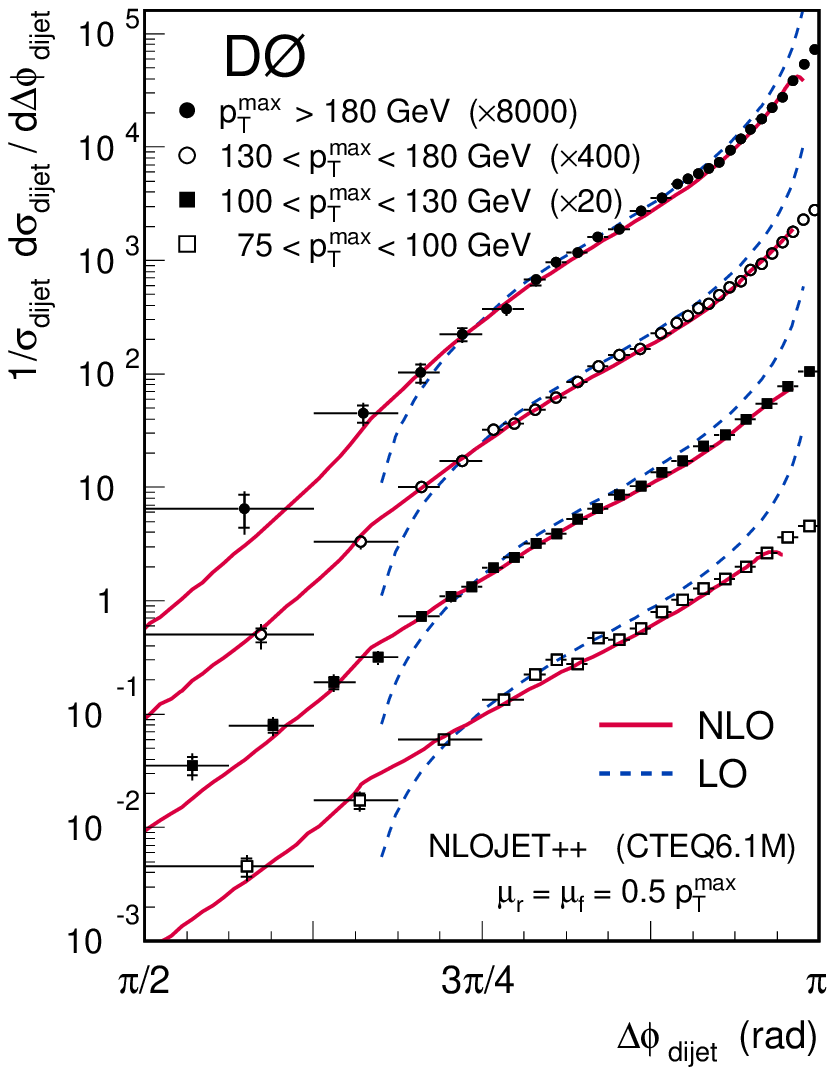}
\caption{\label{f:dphi1} 
The $\Dphi$ distributions in four regions of $\ptmax$.  
Data and predictions with $\ptmax > 100\,\mathrm{GeV}$ are scaled 
by successive factors of 20 for purposes of presentation.
The solid (dashed) lines show the NLO (LO) pQCD predictions.}
\end{minipage}
\hfill
\begin{minipage}[t]{7.5cm}
\includegraphics[scale=0.8]{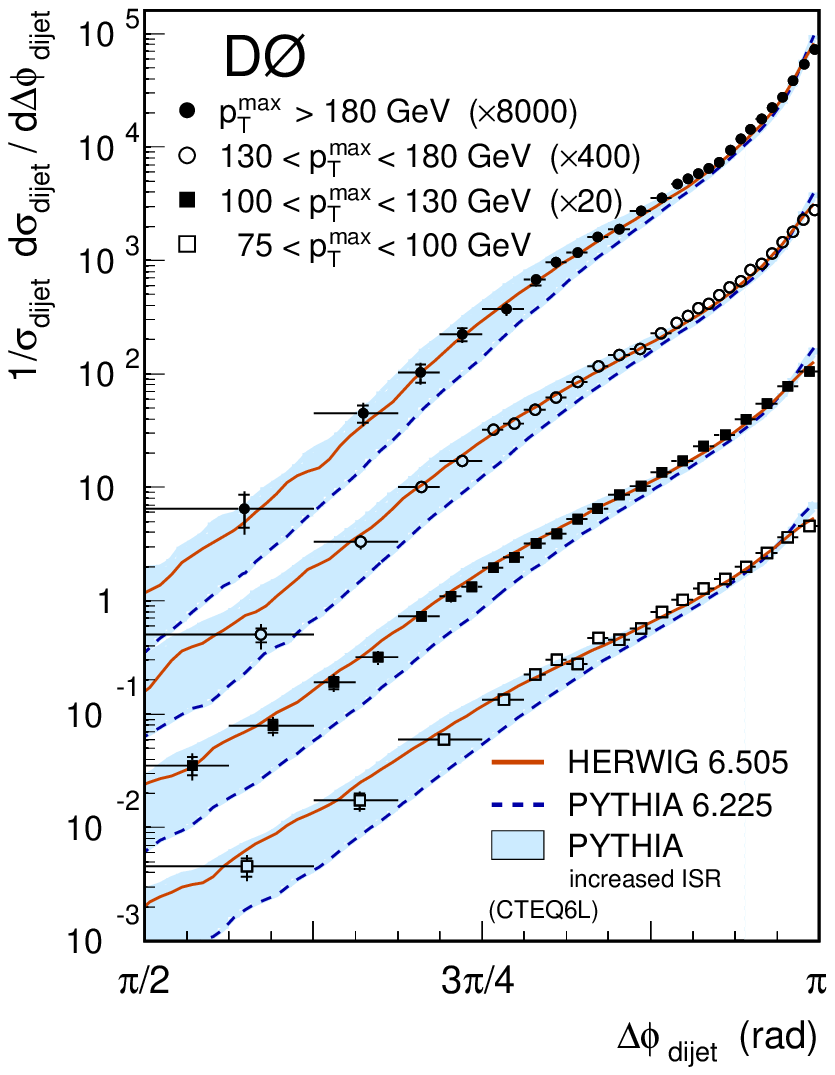}
\caption{\label{fig:data3} 
The $\Dphi$ distributions in different $\ptmax$ ranges.  
Results from {\sc herwig} and {\sc pythia} are overlaid on the data.
Data and predictions with $\ptmax > 100\,\mathrm{GeV}$ are scaled by 
successive factors of 20 for purposes of presentation.}
\end{minipage}
\end{figure}

D\O\ studied radiative processes in QCD by
examining their impact on angular distributions.
D\O\ measured the distribution of azimuthal angle
between two jets with highest $p_{\mathrm{T}}$, 
$\Dphi$~\cite{dphi}.
Second leading jet was required to have $p_{\mathrm{T}}>40\,\mathrm{GeV}$,
and both leading jets were required to have rapidity $|y|<0.5$.
Measurement was performed 
in four bins of leading jet transverse momentum
$\ptmax$. Fully corrected distribution of
dijet azimuthal angle is presented in Fig.~\ref{f:dphi1}.
As the data show, decorrelations increase with decrease of $\ptmax$.

The first non-trivial description of $\Dphi$ distribution is in pQCD
given by the tree level $2\rightarrow3$ parton matrix element.
The limitations of this leading order (LO) prediction
are apparent (see dashed line in  Fig.~\ref{f:dphi1}).
With three partons, it is imposible to produce final state with 
$\Dphi<2\pi/3$. NLO calculations,
obtained with NLOJET++~\cite{nlojet}, provide
much better agreement with data in much wider range of
$\Delta\phi_\mathrm{dijet}$.
However, in the region where $\Dphi\sim\pi$,
any fixed order pQCD calculations become unreliable.
Resummations of soft parton emissions in all orders of pQCD
are needed in order to describe this region properly.


General purpose Monte Carlo (MC) generators, like {\sc Herwig}~\cite{herwig}
or {\sc pythia}~\cite{pythia}, provide such resummations
in the so called leading logarithm
approximation through the developement of parton showers.
Dijet azimuthal decorrelations are then sensitive to the
details of the parton shower mechanism.
Comparison between data and MC generator predictions
is given in Fig.~\ref{fig:data3}.
{\sc herwig} provides good desciption of the data in the
whole range of $\Dphi$, while
{\sc pythia}  gives much smaller
decorrelations than observed in the data.
We found that {\sc pythia} predictions were sensitive to the
parameters of initial-state parton shower (ISR).
Shaded region in Fig.~\ref{fig:data3} indicates the changes
in azimuthal decorrelations as the maximal allowed virtuality
of partons in the shower is increased from its default 
value by factor of four.

\begin{figure}
\begin{minipage}[t]{7.5cm}
\includegraphics[scale=0.4]{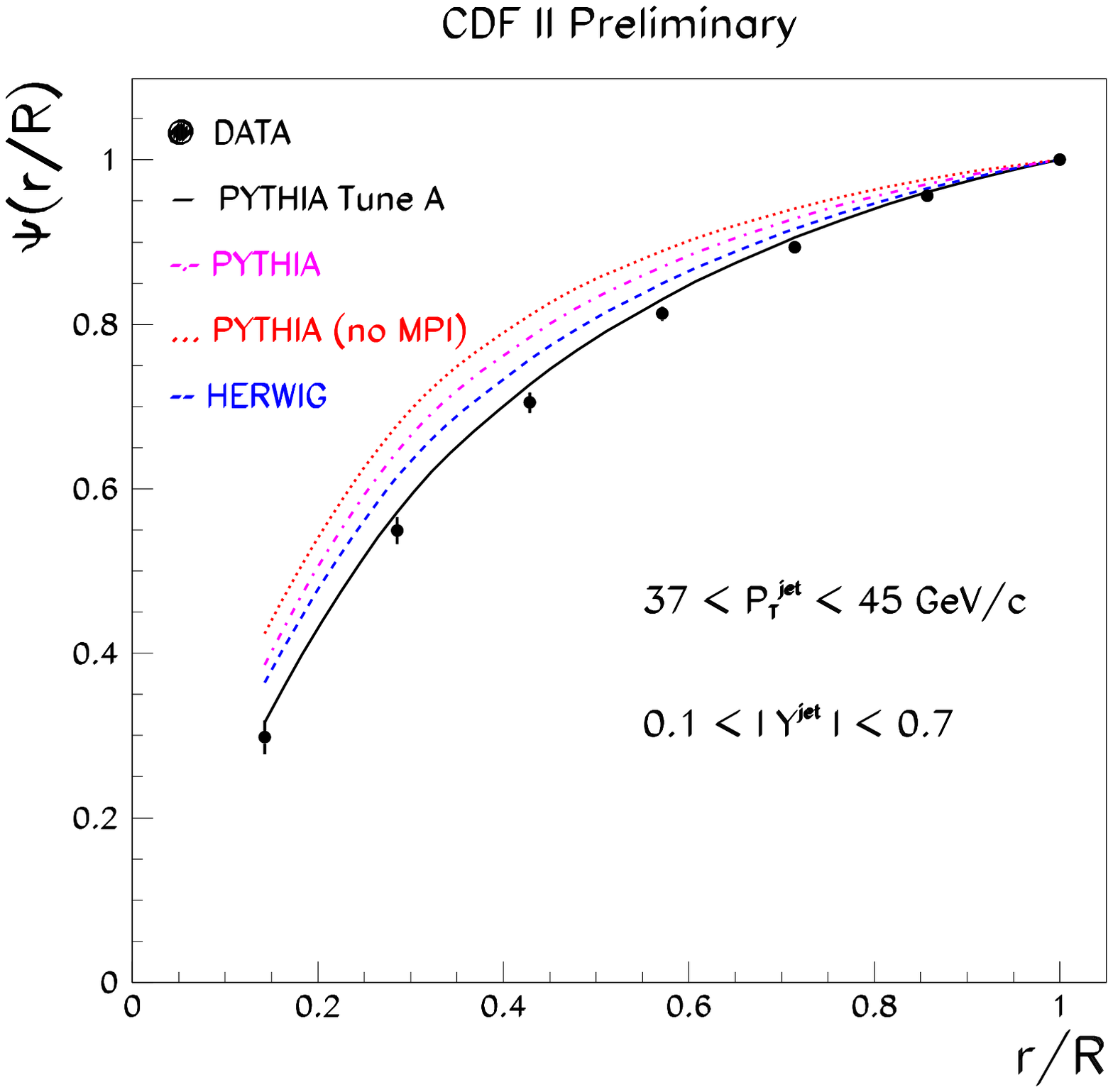}
\caption{\label{f:shape1} Jet shapes in one $p_{\mathrm{T}}$ bin.}
\end{minipage}
\hfill
\begin{minipage}[t]{7.5cm}
\includegraphics[scale=0.4]{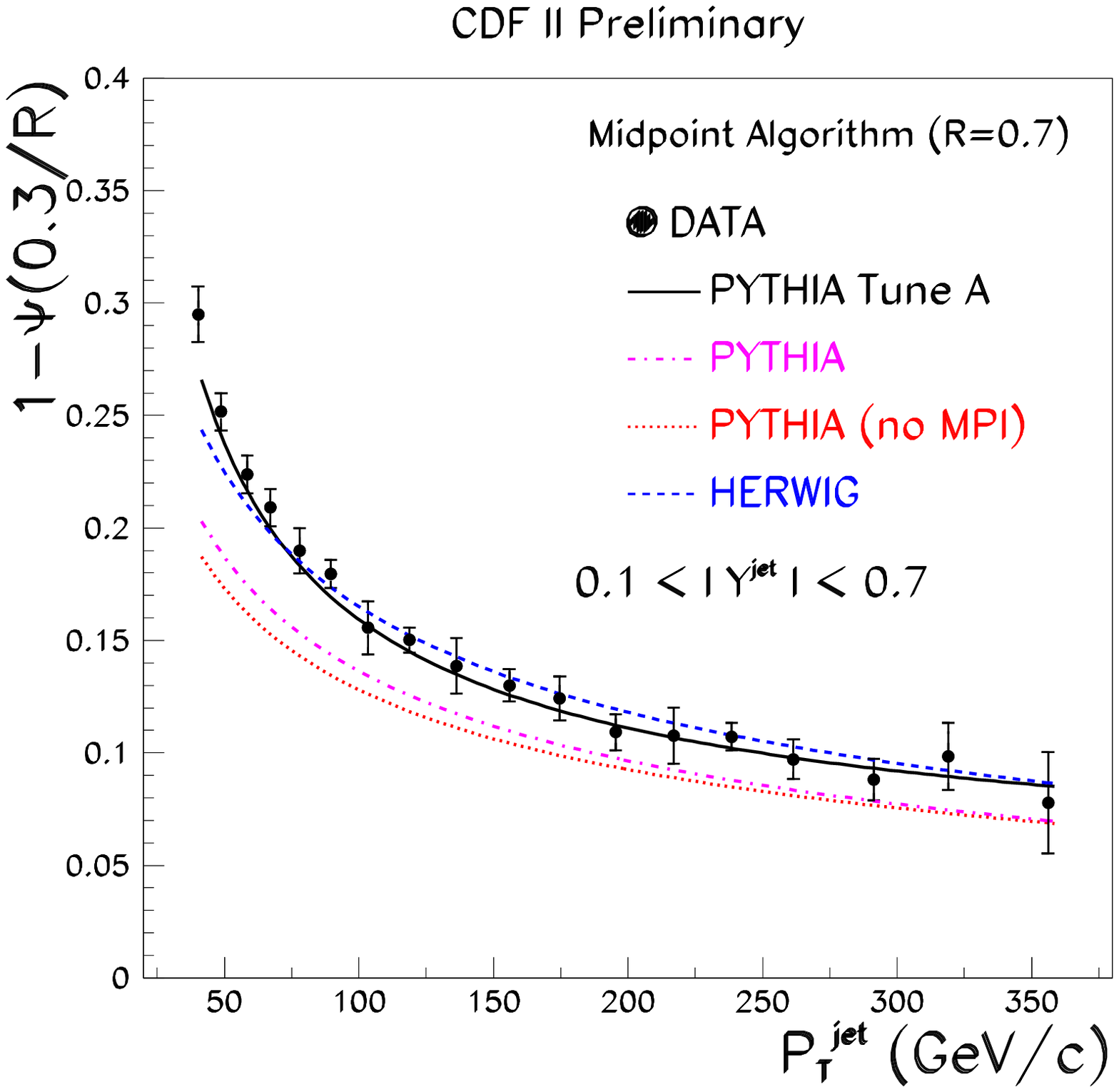}
\caption{\label{f:shape2} Relative jet transverse energy deposited
outside the cone of radius $r=0.3$; dependence on jet $p_\mathrm{T}$.}
\end{minipage}
\end{figure}

Another way how to study the effects of multiparton radiation is to
examine an energy deposition within a jet, so called {\it jet shapes}.
CDF measured averaged deposition of transverse momentum as a function
of distance $r$ from jet exis:
\begin{equation}
\Psi(r) = {1 \over N_{jet}} \sum_{jets} {p_{\mathrm{T}}(0, r) \over
p_{\mathrm{T}}(0, R)}\,,\quad r=\sqrt{\Delta^2\phi + \Delta^2\eta}\,,
\end{equation}
where $p_{\mathrm{T}}(0, r)$ is jet transverse momentum deposited in
cone  with radius~$r$.  Measurement was performed for the central
rapidity jets in wide range of $p_{\mathrm{T}}$.  Jets were
reconstructed with midpoint cone algorithm~\cite{run2cone} with cone
size $R=0.7$.  An example of jet shape measurement in one
low-$p_{\mathrm{T}}$ bin is given in Fig.~\ref{f:shape1}, overall dependence
on jet $p_{\mathrm{T}}$ is then summarized in Fig.~\ref{f:shape2}.
With increasing $p_{\mathrm{T}}$, jets are getting more narrow.
This is due to two reasons: running of $\alpha_{\mathrm{S}}$, and
change of proportion of gluon and quark jets. At low $p_{\mathrm{T}}$,
the sample is dominated by gluon induced jets which are wider than
quark jets, while at high $p_{\mathrm{T}}$, the sample is
dominated by quark jets.

{\sc Herwig}'s and {\sc pythia}'s predictions are compared 
with the data as well in the two figures. {\sc pythia} with default 
setting produces too narrow jets which indicates that there
is not enough radiation in there. This is independent on
{\sc pythia}'s model of soft physics, multi-parton interactions (MPI),
being switched on or off. In Run~I, CDF tuned {\sc pythia}
to their data from studies of soft underlying event.
So called Tune~A gives very good description of jet shapes.
{\sc Herwig} describes data quite well, except the low-$p_{\mathrm{T}}$
region where the jets are narrower than in the data.


To summarize, new results on jet $p_{\mathrm{T}}$ spectra from CDF
and D\O\ experiments has been presented. 
They are consistent with NLO QCD predictions.
Aspects of multi-parton radiation were studied by the two experiments
in the measurement of  decorrelation  in azimuthal angle between the two 
leading jets and in the measurement of jet shapes. 
NLO QCD provides good description of $\Dphi$-distributions except
the region where $\Dphi\sim\pi$.
The results are also useful
for tuning parton shower models in MC generators.


\begin{thebibliography}{99}
\bibitem{cdf} R.~Blair {\it et al.} (CDF II Collaboration), 
              FERMILAB-PUB-96/390-E
\bibitem{d0det} V.~Abazov {\it et al.} (D\O\ Collaboration),
     in preparation for submission to 
       Nucl.~Instrum.~Methods~Phys.~Res.~A;
     T.~LeCompte and H.~T.~Diehl,
       Ann.~Rev.~Nucl.~Part.~Sci. {\bf 50}, 71 (2000);
     S.~Abachi {\it et al.} (D\O\ Collaboration), 
       Nucl.~Instrum.~Methods~Phys.~Res.~A {\bf 338}, 185 (1994).
\bibitem{run2cone} G.~C.~Blazey {\it et al.}, in
     {\sl Proceedings of the Workshop:
     ``QCD and Weak Boson Physics in Run II''},
     edited by U.~Baur, R.~K.~Ellis, and D.~Zeppenfeld, 
     Batavia, Illinois  (2000) p.~47.
     See Section 3.5 for details.
\bibitem{cdfRunIcone} T.~Affolder {\it et al.}  (CDF Collaboration),
  Phys.\ Rev.\ D {\bf 64} (2001) 032001
  [Erratum-ibid.\ D~{\bf 65} (2002) 039903]
\bibitem{cteq6} J.~Pumplin {\it et al.},  JHEP~{\bf 0207}, 12 (2002);
                D.~Stump {\it et al.},   JHEP~{\bf0310}, 046 (2003).
\bibitem{mrst04} 
  A.~D.~Martin, R.~G.~Roberts, W.~J.~Stirling and R.~S.~Thorne,
  Phys.\ Lett.\ B {\bf 604}, 61 (2004)
\bibitem{ellis_soper} S.~D.~Ellis, D.~E.~Soper, Phys.\ Rev.\ D {\bf 48}
                      (1993) 3160.
\bibitem{jetrad} W.~T.~Giele, E.~W.~N.~Glover, Phys.\ Rev.\ D {\bf 46}
                      (1992) 1980.
\bibitem{dphi}
  V.~M.~Abazov {\it et al.}  [D\O\ Collaboration],
  Phys.~Rev.~Lett.~{\bf 94}, 221801 (2005)

\bibitem{nlojet} Z.~Nagy, Phys.~Rev.~Lett.~{\bf 88}, 122003 (2002);
                 Z.~Nagy, Phys.~Rev.~D~{\bf 68}, 094002 (2003).
\bibitem{herwig} G.~Marchesini {\it et al.}, 
          Comp.~Phys.~Comm.~{\bf 67}, 465 (1992); 
          G.~Corcella {\it et al.},  JHEP {\bf 0101}, 010 (2001).
\bibitem{pythia} T.~Sj\"ostrand  {\it et al.},
               Comp.~Phys.~Comm.~{\bf 135}, 238 (2001).

\end{thebibliography}
\end{document}